\documentclass[12pt,a4paper]{article}
\usepackage{cite}
\begin{document}
\title{\bf Does the wavefunction of the universe exist?}
\author{{\bf Thomas Kr\"uger}\thanks{E-mail: t.krueger@phys.uni-paderborn.de}\\Theoretical Physics, Faculty of Science\\University of Paderborn\\Warburger Str. 100, 33098 Paderborn, Germany}
\date{\emph{Dedicated to Prof. Dr. Dieter Schuch, Frankfurt, on the occasion of his 50th birthday.}}
\maketitle
\begin{abstract}
The overwhelming majority of scientists still takes it for granted that classical mechanics (ClM) is nothing but a limiting case of quantum mechanics (QM). Although some physicists restrict this belief to a \emph{generalized} QM as represented, e. g., by the algebra of observables, it will be shown in this contribution that the view of ClM as a mere sub-set of QM is nevertheless unfounded. The usual attempts to derive the laws of ClM from QM are either insufficient or not universally applicable. The transition from traditional to algebraic QM does not add any further insight. It is demonstrated that typical constituents of the classical macroscopic world
\begin{enumerate}
\item cannot be described reasonably in terms of QM and/or
\item do not show up the typical quantum behavior which manifests in the double-slit interference and in the Einstein-Podolsky-Rosen correlations.
\end{enumerate}
Moreover, both attempts to recover ClM from QM and approaches based on vacuum fluctuations are critically inspected, and we arrive at the conclusion: QM does not comprehend ClM, i. e., a wavefunction of the universe does not exist.
\end{abstract}
PACS numbers: 01.70.+w, 03.65.Ta, 03.65.Ud
\newpage
\section{Introduction}
The strategy of Mermin's Ithaca interpretation of quantum mechanics (QM) is "to take the formalism of quantum mechanics as given, and to try to infer from the theory itself what quantum mechanics is trying to tell us about physical reality" \cite{a1}. This approach, however, suffers from the same problem as a lot of other interpretative attempts: The basics are not made clear. A physical theory without a formalism is no physical theory. Indeed the formalism is the heart of any theory, but it is not equivalent to a theory. In order to become a theory any formalism \emph{must} be provided with an interpretation. It is of extreme importance to note that \emph{no} formalism can carry the interpretation in itself. This is evident both from G\"odel's famous theorem \cite{a2} and from a little bit more philosophical considerations: A formalism consists of notions (and the corresponding symbols), relations, and calculation recipes. The recipes are delivered by mathematics, the relations are given by physical experience and/or physical imagination. The notions, however, which form the base of all considerations, are taken from philosophy and our daily experience as human beings. So, in order to establish \emph{any} connection between notions and relations one first of all has to say what each single notion employed shall \emph{mean}, but \emph{meaning} cannot be endowed by a mere formalism. Likewise the cart cannot be put before the horse. Therefore, elements from outside, \emph{not} belonging to the formalism, must be added to it, and in this way meaning is given to the notions. This is both the central and crucial task in any attempt to interpret a formalism, i. e., to create a physical theory \cite{a3}.

The present situation of the interpretation of QM has been reviewed extensively by Lalo\"e \cite{a4} who points out that "the discussions are still lively" which is mainly due to the unsolved question about the 'true nature' of the so-called state vector $ | \Psi \rangle $. One important aspect, however, has not been taken into account: the relation between QM and classical mechanics (ClM). This could be an indication that most members of the physical community take it for granted that ClM is nothing but a limiting case of QM. Said belief can be found in many textbooks (see, e. g., \cite{a5,a6}), and it has been expressed in very plain words by Jauch \cite{a7}: "A classical physical system is $\ldots$ merely a limiting case of a quantal system. It is a system which $\ldots$ behaves in accord with the laws of classical physics. But these laws, as we know, are not exact." Convincing arguments that this belief is nothing but an unfounded opinion are presented in the following sections.

Primas has stated that, in contrast to a \emph{generalized} quantum mechanics, QM in its traditional and well-known formularization is not able to "describe classical systems" \cite{a8}. In his conviction this is a question of the mere formalism and not the fault of any interpretation. The algebra of observables, which represents the most common generalization of QM, is capable to subsume both quantal and classical observables under a common roof indeed, but these are trinkets which do not explain anything! The relation between the classical world and the world of quanta is elucidated not at all if we know that the classical observables form the so-called center of the algebra. It does not help if we use another code for the same old problem although said new code is highly sophisticated. It should have become clear from the first paragraph that a distinct change of the formalism demands for a revision of the interpretation, but this, at least to my knowledge, has not been done so far. The algebra of observables does not yield any answer to the question why most macroscopic entities behave classically while most microscopic entities show up typical quantum behavior. Moreover, what is an observable? There is no precise definition of this notion. But if we employ our rough ideas about the meaning of "observables", we must admit that this does \emph{not} allow for an ontic interpretation of QM. Of course one could refer to statistical operators instead of observables. They at least refer to the entities themselves and not to 'things' we can measure on them, and it can be shown that the set of all statistical operators on a given (finite-dimensional) Hilbert space forms a non-unital $C^*$-algebra with a trivial center. But what is gained thereby for a deeper understanding?

So we are in need of a re-examination of the relation between QM and ClM. It is assumed in general that QM is the more comprehensive theory covering ClM as an asymptotic limiting case. Note that the validity of this assumption is the \emph{condition} that we may speak about the wavefunction of the universe at all. However, in Section 2 it will be shown that this line of reasoning is not tenable. In the following it is argued that a given entity may be considered an element of a certain theory if (and only if) it realizes typical features of the theory. Based on this thesis the possibility of the appearance of \emph{typical} quantum behavior in the realm of human beings as the most prominent constituents of the macroscopic world governed by classical laws is investigated. Section 4 discusses several attempts to \emph{recover} classical from quantum physics, whereas in Section 5 approaches based on vacuum fluctuations are critically inspected. In the concluding section the epistemological consequences are presented.
\section{Classical mechanics - a subset of quantum mechanics?}
Transformed into the language of an ensemble interpretation the usual approach to derive ClM (in its Hamilton-Jacobi form) from QM is as follows: Given a Hilbert space $\mathcal{H}$ with an orthonormal basis set $\{\psi_i\}$ consisting of the eigenfunctions $\psi_i$ of the Hamiltonian $\hat{H} = - \hbar^2/2m \, \Delta + V(r)$ where all spatial coordinates are merged in $r$. Then
\begin{equation}
\rho = \sum_{j,k} c_{jk}(r,t) \, \hat{U}_{jk}
\end{equation}
with $\hat{U}_{jk} = |\psi_j\rangle \langle\psi_k|$ is the most general statistical operator on $\mathcal{H}$, and it is a solution of the equation
\begin{equation}
- \, i \, \hbar \, \frac{\partial \rho}{\partial t} = [\rho,\hat{H}].
\end{equation}
By use of the ansatz
\begin{equation}
c(r,t) = R(r,t) \, \mathrm{e}^{\frac{i}{\hbar} \, S(r,t)}
\end{equation}
where $R$ and $S$ are real functions of $r$ and $t$, and after some lengthy but straightforward manipulations, eq. 2 decays into the two coupled equations
\begin{equation}
\frac{\partial R}{\partial t} = - \frac{1}{2m} \, (2 \, \nabla R \, \nabla S + R \, \Delta S)
\end{equation}
and
\begin{equation}
\frac{\partial S}{\partial t} = - \frac{1}{2m} \, (\nabla S)^2 + E - V + \frac{\hbar^2}{2m} \, \frac{\Delta R}{R}.
\end{equation}
Transformation of (4) yields
\begin{equation}
\frac{\partial R^2}{\partial t} = - \frac{1}{m} \, \nabla(R^2 \, \nabla S)
\end{equation}
which is the well-known continuity equation of ClM \emph{if} $R^2$ is interpreted as a probability density. Except for the expression $\hbar^2/2m \; \Delta R/R$, which is the so-called quantum potential $Q$, eq. 5 is equal to the Hamilton-Jacobi equation (the addition of the $r$-independent energy $E$ to $V$ can be understood as a calibration of the potential). So we \emph{would} have derived the essence of ClM \emph{if} an action in the order of $\hbar^2$ could be neglected. But there are non-exotic cases where the quantum potential is of the form $Q = Q_1(\hbar) + Q_2$ so that the transition $\hbar \rightarrow 0$, even if it \emph{could} be justified, would \emph{not} remove the quantum potential \cite{a9}! Consider, e. g., the time-dependent Schr\"odinger equation of the linear harmonic oscillator:
\begin{equation}
i \hbar \, \frac{\partial \psi}{\partial t} = - \, \frac{\hbar^2}{2 m} \, \frac{\partial^2 \psi}{\partial x^2} + \frac{1}{2} \, m \omega^2 x^2 \psi
\end{equation}
The coherent wavefunction
\begin{eqnarray}
\psi(x,t) &=& \left( \frac{m \omega}{\pi \hbar} \right)^{1/4} \exp \left\{ - \frac{m \omega}{2 \hbar} \, (x - \langle x(t) \rangle)^2 \right\} \nonumber \\ & & \times \exp \left\{ \frac{i}{\hbar} \left[ \langle p(t) \rangle \, x - \frac{\hbar \omega}{2} \, t - \frac{\langle x(t) \rangle \, \langle p(t) \rangle}{2} \right] \right\}
\end{eqnarray}
is a solution of (7). $\langle x(t) \rangle$ and $\langle p(t) \rangle$ are the expectation values of position and momentum, respectively, at the time $t$, and it is well known that they follow classical trajectories. By comparing (8) and (3) it is seen immediately that
\begin{equation}
R(x,t) = \left( \frac{m \omega}{\pi \hbar} \right)^{1/4} \exp \left\{ - \frac{m \omega}{2 \hbar} \, (x - \langle x(t) \rangle)^2 \right\}.
\end{equation}
The function $S(x,t)$ is then given by the term $[ \ldots ]$ in (8). We form the second derivative of $R$ with respect to $x$ and obtain the quantum potential
\begin{equation}
Q(x,t) = \frac{\hbar \omega}{2} - \frac{m \omega^2}{2} \, (x - \langle x(t) \rangle)^2,
\end{equation}
which, however, will \emph{not} tend to zero if $\hbar \rightarrow 0$, i. e., even in this simple and non-exotic case the neglect of $\hbar$ will not remove the non-classical quantum potential.
Up to now, however, this fact is still ignored quite frequently. Michael Berry, e. g., stated recently \cite{a10} that the "classical world is the limit of the quantum world when Planck's constant \emph{h} is inappreciable."

Furthermore, the classical limit, i. e., the 'point' where ClM emerges from QM, is usually discussed in terms of Ehrenfest's theorem which states that, if only a wave packet is sufficiently narrow, then its mean position will follow a classical trajectory. But Ballentine et al. have proven that Ehrenfest's theorem is neither a necessary nor a sufficient condition to characterize the quantal$\rightarrow$classical transition \cite{a11}.

The famous superposition principle states that, given two statistical operators $\rho_1$ and $\rho_2$ on $\mathcal{H}$, then also the convex sum
\begin{equation}
\rho = w \rho_1 + (1 - w) \rho_2
\end{equation}
with $0 < w < 1$ is a statistical operator on $\mathcal{H}$. This principle, which is a consequence of the linearity of (2), has no counterpart in ClM. Recall that the dynamics of $\rho$ is governed by the quantum analogue of the classical Liouville equation, $\mathrm{i} \partial \rho_{classical}/\partial t = \hat{L} \rho_{classical}$, which is linear as well. The stationary probability distributions are eigenfunctions of the Liouville operator $\hat{L}$ to the eigenvalue 0. Now the ergodic hypothesis states that there is only \emph{one} corresponding eigenfunction which is completely determined by the ensemble's Hamilton function. With only one $\rho_{classical}$, however, a superposition is impossible. 

These three facts shed considerable doubt on the belief that ClM is a mere subset of QM. This situation is illustrated nicely by using QM to describe features of the solar system. If QM were universally valid, then all details of the motion of the earth around the sun should be given by the solution of the eigenvalue problem of the Hamiltonian
\begin{equation}
\hat{H}^{solar} = - \frac{\hbar^2}{2 M_1} \, \Delta(x,y,z) - \frac{G M_1 M_2}{r}
\end{equation}
where, for the sake of simplicity, the origin of the coordinate system has been identified with the sun's center of mass. $G$ is the gravitational constant, $M_1$ the mass of the earth, and $M_2$ the mass of the sun. By transformation of the Cartesian coordinates into spherical coordinates we obtain
\begin{equation}
\left( - \frac{\hbar^2}{2 M_1} \left( \frac{2}{r} \, \frac{\partial}{\partial r} + \frac{\partial^2}{\partial r^2} + \frac{1}{r^2} \left[ - \frac{1}{\hbar^2} \, \hat{L}^2 \right] \right) - \frac{G M_1 M_2}{r} \right) \, |\Psi\rangle = E \, |\Psi\rangle.
\end{equation}
With $|\Psi\rangle = R \, '(r) Y_{l,m}(\theta,\varphi)$ and $L^2 = \hbar^2 \, l (l + 1) \; (l \in \mathbf{N_0})$ eq. 13 can be transformed into
\begin{equation}
\left( - \frac{\hbar^2}{2 M_1} \left( \frac{2}{r} \, \frac{d}{d r} + \frac{d^2}{d r^2} \right) + \frac{\hbar^2 \, l (l + 1)}{2 M_1 r^2} - \frac{G M_1 M_2}{r} - E \right) \, R \, '(r) = 0.
\end{equation}
We abbreviate
\begin{equation}
\frac{G M_1^2 M_2}{\hbar^2} := \alpha
\end{equation}
and
\begin{equation}
\frac{M_1 E}{\hbar^2} := \varepsilon,
\end{equation}
and obtain after some lengthy but straightforward manipulations in analogy to the treatment of the hydrogen atom the well-known Laguerre equation:
\begin{equation}
x \, w'' + (2(l + 1) - x) \, w' + \left( \frac{1}{c} - (l + 1) \right) w = 0
\end{equation}
This equation has solutions if and only if $1/c = n \; (n \in \mathbf{N})$ and $n \ge l + 1$. From this we obtain the condition $\alpha^2/(2|\varepsilon|) = n^2$ which yields a formula for the energy eigenvalues:
\begin{equation}
E_n = - \frac{G^2 M_1^3 M_2^2}{2 \hbar^2} \times \frac{1}{n^2}
\end{equation}
By inserting the respective values of the constants we finally end at
\begin{equation}
E_n = - 1.70 \times 10^{182} \; \mathrm{J} \; \times \frac{1}{n^2}.
\end{equation}
If this application of QM to the solar system shall make any sense at all, then the quantum mechanical total energy must be at least approximately equal to the total energy obtained by ClM. In the latter case it is easy to see that the sum of kinetic and potential energy amounts to $7.96 \times 10^{33} \; \mathrm{J}$. So, to establish equivalence between $E_{quantal}$ and $E_{classical}$, $n$ must be equal to about $10^{74}$! Such quantum numbers, although not impossible in principle, are \emph{completely meaningless} because the functions $w$ satisfying (17) are essentially the associated Laguerre polynomials, and it will \emph{never} be possible to evaluate polynomials with terms up to the $10^{74}$th power! But this is a point of minor importance only. Instead we have to ask what a quantum number of $n = 10^{74}$ actually means, and the answer is quite simple: Obviously our solar system is not in its ground state but in the $10^{74}$th \emph{excited} state! This, however, is really hard to believe, because coupling to a gravitational field would lead to spontaneous emission of energy yielding finally the solar system in its ground state, but one could assume that the transition from $n = 10^{74}$ to $n = 1$ takes such a long time that the stability of the solar system is guaranteed for all practical purposes. Let us see whether this assumption is correct.

Coupling to a gravitational field causes the decay of the initial wavefunction $\Psi_n = \psi_n \, \exp(i E_n t/\hbar)$ according to
\begin{equation}
\Psi_n \rightarrow c \Psi_n + \sum_{k=1}^{\infty} c_k(t) \Psi_k
\end{equation}
with $c_k(0) = 0 \; \forall \; k$ and $|c|^2 = 1$. The functions $\psi_k$ shall be eigenfunctions of the undisturbed Hamiltonian $\hat{H}_0$. Insertion into the time-dependent Schr\"odinger equation leads to the following differential equation for the coefficients $c_k$:
\begin{equation}
\frac{\hbar}{i} \, \sum_k \frac{d c_k}{d t} \, \psi_k \, \mathrm{e}^{- i E_k t / \hbar} = - \hat{H}^{\prime} c \psi_n \, \mathrm{e}^{- i E_n t / \hbar} - \sum_k c_k \hat{H}^{\prime} \psi_k \, \mathrm{e}^{- i E_k t / \hbar}
\end{equation}
We multiply (21) from the left by $c^* \psi_n^* \, \exp(i E_n t / \hbar)$ and integrate over the spatial coordinates.
\begin{equation}
\Rightarrow \frac{\hbar}{i} \, \frac{d c_n}{d t} = - \, \langle \psi_n | \hat{H}^{\prime} | \psi_n \rangle - \sum_k c_k \, \langle \psi_n | \hat{H}^{\prime} | \psi_k \rangle \, \mathrm{e}^{- i \omega_{kn} t}
\end{equation}
where $\omega_{kn} = (E_k - E_n)/\hbar$.
To get an idea of the orders of magnitude we are speaking about let us assume, for the moment, that the external gravitational field is constant. [A more detailed treatment can be found in Appendix A.] We choose it to be the gravitational field of jupiter at the center of our coordinate system, i. e.,
\begin{equation}
\hat{H}^{\prime} \equiv H^{\prime} = \frac{G M_1 M_3}{r_{SJ}}
\end{equation}
where $M_3$ is the mass of jupiter and $r_{SJ}$ is its mean distance from the sun located at the center of our coordinate system. Since $H^{\prime}$ is constant, (22) can be simplified significantly, and we obtain
\begin{eqnarray}
\frac{\hbar}{i} \, \frac{d c_n}{d t} &=& - H^{\prime} \langle \psi_n | \psi_n \rangle - \sum_k c_k \, H^{\prime} \underbrace{\langle \psi_n | \psi_k \rangle}_{= \delta_{nk}} \, \mathrm{e}^{i \omega_{nk} t} \nonumber \\ &=& - H^{\prime} \, (1 + c_n).
\end{eqnarray}
The solution of this differential equation is
\begin{equation}
c_n(t) = \mathrm{e}^{- i H^{\prime} t / \hbar} - 1.
\end{equation}
The decay of the initial wavefunction is complete if the superposition (20) does not contain any fraction of $\Psi_n$ any more. A necessary condition for this is $|c_n(t)|^2 = 1$. For the time required for this process we then obtain
\begin{equation}
t_d = \frac{\hbar}{H^{\prime}} \, \arccos \frac{1}{2}
\end{equation}
which yields a value of $1.14 \times 10^{-64}$ s only, i. e., the solar system as we know it would have collapsed already at its very beginning! Note that the coefficient $c_n(t_d)$ amounts to $-1/2 - i \surd 3/2$ which exactly cancels $c$ in (20) if we choose $c = 1/2 + i \surd 3/2$. [A more sophisticated analysis of the solar system's stability taking into account the external perturbation in detail can be found in Appendix A. It leads, however, to the same conclusion.]

Furthermore, the ground state, characterized by $n = 1$ and $l = 0$, possesses a radial density of $4 \rho^2 \exp(- 2 \rho)$ which has its maximum at a distance $r_m = 2.34 \times 10^{- 138}$ m from the origin of the coordinate system, i. e., the most possible distance between sun and earth in the ground state of this system amounts to $r_m$ - which by sure would be a very uncomfortable place to stay. Out of these reasons QM must be considered \emph{totally inadequate} to replace ClM in the description of features of our solar system.

One could object, using an example given by Berry \cite{a10}, that, in certain cases, decoherence reinstates classicality so that an inadequate quantum description can be corrected by considering a respective \emph{open} quantum system. But why should one even start with a quantum description if the classical description must be reconstructed cumbersomely anyway?
\section{Quantum behavior in the macroscopic \\ world}
\begin{quote}
Thesis: \emph{If, under suited conditions, a given entity shows a behavior which is typical for a certain theory, then said entity is an element of this theory.}
\end{quote}
Superconductivity is a phenomenon which can be understood only by means of QM, and in view of the fact that modern particle accelerators use superconducting coils with a volume of several m$^3$, one could be tempted to assign these coils as massive 'things' of macroscopic dimensions to the realm of QM. Note, however, that superconductivity gets totally lost if the coils are warmed up to room temperature. In this case nobody would speak about a coil as a quantum object. This temperature-dependency of the assignment seems to be mind-puzzling, but of course one could reply that there is no mystery at all because every information we obtain about a system depends on the way we ask, i. e., the information is \emph{contextual}. So we must ask in a proper way.

Superconductivity, however, is not a feature of QM. It is not an intrinsic and unavoidable consequence of the theory, becausse no axiom or postulate states that all quantum objects \emph{must} be superconducting if the conditions are chosen properly. QM merely \emph{allows} for this phenomenon. Therefore superconductivity can not be considered a \emph{typical} quantum phenomenon. It is a quantum phenomenon, but not a typical one, and in consequence the question about the theory assignment of coils can not be answered yet.

In my opinion there are only two phenomenological aspects where QM differs decisively from its classical counterpart: wave-particle duality and the Einstein-Podolsky-Rosen (EPR) correlations. Neither of the two is known in classical theories, but according to QM \emph{all} of its elements should be able to realize these aspects under suited circumstances. So, if QM is the comprehensive theory, then it must be possible - at least in principle - to demonstrate the existence of both wave-particle duality and EPR correlations with the aid of entities which are typically assigned to ClM. In the following we will therefore discuss these typical features of QM in terms of human beings.
\subsection{Wave-particle duality}
This phenomenon is demonstrated best in the famous double-slit experiment. Assume that a source emits a vast amount of entities, one after the other, which can be considered to be particles. These particles hit a double-slit arrangement where $b$, the width of each slit, is much smaller than the distance $c$ between the slits. A registration screen is placed at a distance $d$ behind the double-slit. It is well known that, if this experiment is performed with neutrons, e. g., an interference pattern is recorded which can be explained if and only if the neutrons are assumed to show up a wave-like behavior.

Zeilinger and coworkers have replaced the ensemble of neutrons by an ensemble of C$_{60}$ fullerenes sublimated in an oven \cite{a12}. The emitted molecular beam traversed a grating with period $s = b + c = 100$ nm, and impinged on a spatially resolving detector placed 1.25 m behind the grating. Zeilinger and coworkers succeeded, for the first time, to observe an interference pattern with 'real' molecules. The distance $x_1$ between the central peak and the next maximum amounted to about 30 $\mu$m.

From standard diffraction theory we know that
\begin{equation}
\frac{\lambda}{s} = \sin \left(\arctan \frac{x_1}{d}\right),
\end{equation}
i. e., the achievable resolution $x_1$ depends on the distance between grating (or double-slit) and detector, on the grating period $s$, and on the wavelength $\lambda$ of the incoming entities. Now assume that this experiment is repeated with an ensemble of physicists with a mass $m$ of 90 kg and jogging towards the grating with a velocity $v$ of 10 km/h. Their corresponding de Broglie wavelength $\lambda = h/mv$ amounts to $2.65 \times 10^{-36}$ m. Let us hope that we are in possession of an extremely sensitive detector allowing for a spatial resolution $x_1$ of 1 nm. Then, even if the detector would be placed 1 km behind the grating, the period $s$ necessary to produce an interference pattern would amount to $2.65 \times 10^{-24}$ m which is \emph{14 orders of magnitude smaller} than typical atomic diameters! Since every slit must have a minimum width of one atom layer, it will never be possible to produce a grating for this experiment. But note that this is \emph{not} a technical problem! Slits with a dimension much smaller than atomic nuclei do not exist in principle, i. e., on this length scale the notion "slit" does not have any meaning at all!

The last years have brought considerable progress in atom interferometry with gratings made of \emph{light}. Their period $s$ amounts to one half of the wavelength $\lambda_{light}$ of the radiation employed to generate it. Using the example given above it is seen immediately that $\lambda_{light}$ must be equal to $5.30 \times 10^{-24}$ m corresponding to an energy of $2.37 \times 10^{17}$ eV or a temperature of $2.71 \times 10^{21}$ K. It is hard to believe that the universe will ever be able to provide us with a radiation source of the required power. So, having shown that there is no chance at all to \emph{detect} wave-particle duality in the case of human beings, why should one believe that there \emph{is} wave-particle duality in the macrocosmos at all? 
\subsection{EPR correlations}
A given quantum entity shall decay at a time $t_0$ into two parts called U$_i$ and V$_i$, respectively. Suppose that U$_i$ and V$_i$ fly away from one another. An experimental setup shall separate U$_i$ and V$_i$ spatially so that they cannot interact any more by means of any known physical principle. U$_i$ (V$_i$) shall impinge on apparatus A (B). A and B are of the same kind. At a time $t_1$ sufficiently larger than $t_0$ a property type as, e. g., spin component or polarization axis shall be measured on both U$_i$ and V$_i$. Let us choose this property type E to be dichotomic in such a way that both the property (the numerical value of the property type) $E$(U$_i$) and $E$(V$_i$) shall be equal to $\pm 1$ at certain experimental parameters $\vec{a}$ and $\vec{b}$, respectively, which are assumed to determine the actual internal structure of the two coupled apparatuses A and B. $\vec{a}$ and $\vec{b}$ are, e. g., the unit vectors defined by either the directions of the inhomogeneous magnetic field in two Stern-Gerlach devices or the axes of two polarizers. Then the outcome of one single run is given by
\begin{equation}
O_i(\vec{a},\vec{b}) = E(\vec{a};\mathrm{U}_i) \, E(\vec{b};\mathrm{V}_i).
\end{equation}
Suppose further that a sufficient number of single runs has been done. The final result $O(\vec{a},\vec{b})$ is the mean of all single outcomes.

We associate the ensemble of produced or emitted entity pairs (U$_i$,V$_i$) with a statistical operator $\rho$ on a four-dimensional Hilbert space $\mathcal{H}$ which is the direct product of the two orthogonal Hilbert spaces $\mathcal{H}_U$ and $\mathcal{H}_V$. $\mathcal{H}_U$ ($\mathcal{H}_V$) is connected to the physical space wherein E is measured on U$_i$ (V$_i$) using A (B). Let $\{|\alpha_1\rangle,|\alpha_2\rangle\}$ be an orthonormal basis of $\mathcal{H}_U$ and $\{|\beta_1\rangle,|\beta_2\rangle\}$ an orthonormal basis of $\mathcal{H}_V$.

The apparatus to measure E on U$_i$ (V$_i$) shall be represented by the self-adjoint operator $\hat{A}$ ($\hat{B}$) acting on $\mathcal{H}_U$ ($\mathcal{H}_V$). $\hat{A}$ and $\hat{B}$ are determined by the experimental parameters $\vec{a}$ and $\vec{b}$, respectively, which are mentioned above. The combination of the two apparatuses to perform measurements of E on both U$_i$ and V$_i$ yielding $O(\vec{a},\vec{b})$ has to be represented by the direct product of $\hat{A}$ and $\hat{B}$ according to
\begin{equation}
\hat{P}(\vec{a},\vec{b}) = \hat{A} \otimes \hat{1}_V \times \hat{1}_U \otimes \hat{B} = \hat{A} \otimes \hat{B}.
\end{equation}
Apparatus B differs from A insofar as $\vec{b} \ne \vec{a}$. Let $\vec{a}$ be a principal axis in the laboratory coordinate system. Then $\hat{B}$ emerges from $\hat{A}$ by a rotation around the angle $\chi$ between the two vectors $\vec{a}$ and $\vec{b}$. In complete analogy we define two further operators, $\hat{A}'$ and $\hat{B}'$, where $\hat{A}'$ represents apparatus A rotated with respect to its first position by an angle $\varphi$. $\hat{B}'$ stands for B rotated with respect to $\vec{a}$ by an angle $\psi$. Note that $\hat{A}$ and $\hat{A'}$ as well as $\hat{B}$ and $\hat{B'}$ are in general \emph{non}-commuting operators.

With the aid of these four operators we can simulate four different experimental arrangements. The corresponding operators of the joint measurements (the coincidence operators) are $\hat{P}(\vec{a},\vec{b})$, $\hat{P}(\vec{a},\vec{b}')$, $\hat{P}(\vec{a}',\vec{b})$, and $\hat{P}(\vec{a}',\vec{b}')$. Each of these experiments, if actually performed, leads to a final result $O$ from which the correlation function $\Delta$ can be determined:
\begin{equation}
\Delta = \left|O(\vec{a},\vec{b}) - O(\vec{a},\vec{b}')\right| + \left|O(\vec{a}',\vec{b}) + O(\vec{a}',\vec{b}')\right|
\end{equation}
On the other hand, $\Delta$ can be \emph{calculated} as well because in QM the final outcome of a measurement on an ensemble of entities is given by
\begin{equation}
O(\vec{a},\vec{b}) = \mathrm{Tr}\left(\hat{P}(\vec{a},\vec{b}) \: \rho\right).
\end{equation}
In this way $\Delta$ becomes dependent on the choice of the statistical operator.

The set of all statistical operators on $\mathcal{H}$ consists of two mutually exclusive subsets, because a statistical operator can be either separable or not. We call $\rho$ K-separable if (and only if)
\begin{equation}
\rho = \rho_U \otimes \rho_V
\end{equation}
where $\rho_U$ is acting on $\mathcal{H}_U$ and $\rho_V$ on $\mathcal{H}_V$. This definition is a simplification of the usual one which offers a couple of advantages explained in \cite{a3}. A detailed comparison of the two definitions can be found in \cite{a13}.

Now let us turn to the 'real' world. Assume that there is an ensemble of physicists and let each member of the ensemble have one of two 'orthogonal' political convictions, say, R and L, respectively. Then the ensemble is represented by a statistical operator $\rho_U$ on the two-dimensional Hilbert space $\mathcal{H}_U$ with the orthonormal basis functions $|\alpha_R\rangle$ and $|\alpha_L\rangle$ according to
\begin{equation}
\rho_U = \sum_{i,j} u_{ij} \, \hat{U}_{ij}
\end{equation}
with $\hat{U}_{ij} = |\alpha_i\rangle \langle \alpha_j|$. Now assume that there is a second ensemble of physicists, represented by $\rho_V$, whose political opinions are also either R or L. At a time $t_1$ we allow for a contact between these two ensembles. So the physicists will exchange views and in consequence some of them will change their conviction or become undecided. At this stage we exclude the possibility that members of the same ensemble are engaged in a discussion. Only inter-ensemble contacts are admitted. After some time interval $\Delta t$ we stop the interaction and re-separate the two ensembles, but while at $t < t_1$ the \emph{total} ensemble $\{\{\mathrm{U}_i\},\{\mathrm{V}_i\}\}$ is represented by $\rho_U \otimes \rho_V$, it now should be represented by the most general statistical operator on the product Hilbert space which is given by
\begin{equation}
\rho_{after} = \sum_{i,j,k,l} c_{ij,kl} \, \hat{U}_{ij} \otimes \hat{V}_{kl}.
\end{equation}
In general $\rho_{after}$ will be non-K-separable although $\rho_{before}$ (see eq. 32) has been K-separable. This is the effect of the entangling interaction between the two ensembles. A similar process has been used in a recent proposal of an EPR-type experiment with molecules \cite{a14}.

At a time $t_2 > t_1 + \Delta t$ the members of both ensembles have to undergo a lie detector test one by one. In this first measurement series detector A shall be represented, in matrix form, by
\begin{displaymath}
\sigma_3 = \left( \begin{array}{cc} 1 & 0 \\ 0 & -1 \end{array} \right)
\end{displaymath}
and B by $- \sigma_3$ so that, if A detects R and B detects L, the coincidence result will be $(+1) \times (+1) = 1$. Now recall that \emph{four} measurement series are needed in order to yield $\Delta$. So we define
\begin{equation}
\hat{A}' = \sigma_1 \; \; \mathrm{and} \; \; \hat{B}' = - \sigma_1.
\end{equation}
This choice represents a device setting suited to detect total undecidedness, i. e., the situation where the political opinion of a member of one of the ensembles is a 50:50 mixture of R and L. Note that $\hat{A}' \; (\hat{B}')$ is the result of a rotation of $\hat{A} \; (\hat{B})$ by 90$^{\circ}$. The operators $\hat{A}$ and $\hat{A}'$ ($\hat{B}$ and $\hat{B}'$) are maximally non-commuting in the sense that the absolute value of the determinant of the commutator attains its maximum.

Using the non-K-separable operator $\rho_{after}$ defined by eq. 34 we obtain after a lengthy but straightforward calculation
\begin{eqnarray}
\Delta_{n-s} &=& \left| - c_{11,11} + c_{11,22} + c_{22,11} - c_{22,22} + c_{11,12} + c_{11,21} - c_{22,12} - c_{22,21} \right| \nonumber \\ & & + \, \left| - c_{12,11} - c_{21,11} + c_{12,22} + c_{21,22} - c_{12,12} - c_{12,21} - c_{21,12} - c_{21,21} \right|. \nonumber \\ & &
\end{eqnarray}
The four eigenvalues $\lambda_n$ of $\rho_{after}$ are functions of the coefficients $c_{ij,kl}$. The requirement $\lambda_n \ge 0 \; \forall \; n$ imposes four conditions on the non-diagonal coefficients so that there is a certain arbitrariness in fixing them. We therefore assume, e. g., that only the coefficients forming the anti-diagonal are different from zero, i. e., $\rho_{after}$ corresponds to the matrix
\begin{equation}
\left( \begin{array}{cccc} c_{11,11} & 0 & 0 & x_1 + \mathrm{i} y_1 \\ 0 & c_{11,22} & x_2 + \mathrm{i} y_2 & 0 \\ 0 & x_2 - \mathrm{i} y_2 & c_{22,11} & 0 \\ x_1 - \mathrm{i} y_1 & 0 & 0 & c_{22,22} \end{array} \right).
\end{equation}
In consequence (36) is simplified significantly and we obtain
\begin{equation}
\Delta_{n-s} = \left|- c_{11,11} + c_{11,22} + c_{22,11} - c_{22,22}\right| + 2 \, |- x_1 - x_2 |.
\end{equation}
For given diagonal coefficients this expression is maximized if the second term becomes maximal. The eigenvalues of $\rho_{after}$ are
\begin{equation}
\lambda_{1,2} = \frac{1}{2} \, \left\{ C \pm \sqrt{C^2 - 4 \left( c_{11,11} C - c_{11,11}^2 - (x_1^2 + y_1^2) \right)} \right\} 
\end{equation}
with
\begin{equation}
C = 1 - c_{11,22} - c_{22,11}
\end{equation}
and
\begin{equation}
\lambda_{3,4} = \frac{1}{2} \, \left\{ c_{11,22} + c_{22,11} \pm \sqrt{(c_{11,22} - c_{22,11})^2 + 4 \, (x_2^2 + y_2^2)} \right\}.
\end{equation}
The conditions $\lambda_{2,4} \ge 0$ yield upper bounds for $x_1^2 + y_1^2$ and $x_2^2 + y_2^2$, respectively, but since $y_{1,2}$ do not appear in (38) they can be set equal to zero in order to allow for maximum values of $x_1^2$ and $x_2^2$. We then arrive at
\begin{equation}
x_1^2 \le c_{11,11} c_{22,22} \; \; \mathrm{and} \; \; x_2^2 \le c_{11,22} c_{22,11}
\end{equation}
and obtain by insertion into (38)
\begin{equation}
\Delta_{n-s} \le \left|- c_{11,11} + c_{11,22} + c_{22,11} - c_{22,22}\right| + 2 \, \left( \sqrt{c_{11,11} c_{22,22}} + \sqrt{c_{11,22} c_{22,11}} \right).
\end{equation}
Numerical evaluation yields the final result $\Delta_{n-s} \le 2$.

Now reconsider the experimental situation. If the two formerly independent ensembles are separated after a certain interaction period, then their future fates will be independent of one another. Future interaction of \{U$_i$\} with another ensemble \{W$_i$\} or with TV spots and newspapers may change the political opinion of some of the members of \{U$_i$\}, but this change will not at all influence the V$_i$. Therefore the two ensembles \{U$_i$\} and \{V$_i$\} \emph{may} be considered totally \emph{independent} so that this system could be a realization of EPR's 1935 principle of separability. In consequence \{\{U$_i$\},\{V$_i$\}\} might also be represented by the separable operator $\rho$ defined by (32). But which one of the two possible statistical operators is the proper representation of the total ensemble? This question can be answered only if $\rho$ and $\rho_{after}$ lead to different predictions regarding $\Delta$ which can be checked experimentally.

With $\rho$ we obtain the correlation function
\begin{eqnarray}
\Delta_s &=& \left|(u_{11} - u_{22}) \, (- v_{11} + v_{12} + v_{21} + v_{22})\right| \nonumber \\ & & + \, \left|(u_{12} + u_{21}) \, (- v_{11} - v_{12} - v_{21} + v_{22})\right|.
\end{eqnarray}
$u_{12}$ ($v_{12}$) can be written as $x_1 + \mathrm{i} y_1$ ($x_1 + \mathrm{i} y_1$), and the non-negativity of $\rho_U$ and $\rho_V$ is guaranteed if $x_1^2 \le u_{11} u_{22}$ and $x_2^2 \le v_{11} v_{22}$, respectively. So (44) reduces to
\begin{eqnarray}
\Delta_s &\le& \left|(u_{11} - u_{22}) \, (- v_{11} + 2 \sqrt{v_{11} v_{22}} + v_{22})\right| \nonumber \\ & & + \, \left|2 \sqrt{u_{11} u_{22}} \, (- v_{11} - 2 \sqrt{v_{11} v_{22}} + v_{22})\right|,
\end{eqnarray}
and numerical evaluation yields $\Delta_s \le \sqrt{2}$.

The difference between the upper bounds of $\Delta_s$ and $\Delta_{n-s}$ is significant. $\Delta_{n-s}$, however, does \emph{not} violate Bell's inequality. Moreover, if we replace the matrix representation (37) of $\rho_{after}$ by, e. g.,
\begin{equation}
\left( \begin{array}{cccc} c_{11,11} & 0 & x_1 + \mathrm{i} y_1 & 0 \\ 0 & c_{11,22} & 0 & x_2 + \mathrm{i} y_2 \\ x_1 - \mathrm{i} y_1 & 0 & c_{22,11} & 0 \\ 0 & x_2 - \mathrm{i} y_2 & 0 & c_{22,22} \end{array} \right),
\end{equation}
which is an admissible choice as well, and proceed in the same way as described above, we obtain the inequality $\Delta_{n-s} \le 1.40$! 
In view of the fact that one should speak about EPR correlations only if Bell's inequality is violated, we cannot make any unambiguous statement regarding the correlation of the two ensembles in question unless the entangling interaction is understood in detail. This, however, does not seem to be possible at all. Therefore it is at least very doubtful whether EPR correlations appear in the realm of typical classical entities.

\section{Recovering classical physics}
The arguments of the preceding sections by no means justify to consider QM the more comprehensive and fundamental theory. Nevertheless it could be possible that, in the case of very large systems or entities, a quantum mechanics can be found which allows for a description which is isomorphic to what would be obtained by ClM. Corresponding attempts are subsumed under the notion "recovering classical physics." There are several justifications for this program. Joos and Zeh, on the one hand, have shown that classical properties may emerge through interaction of a quantum system with a large environment \cite{a15}. Their work, however, is based on the belief that QM is universally valid - which must be put into doubt because of the arguments given above. On the other hand there is a lot of evidence that the superposition principle looses its validity by decoherence through coupling to an environment (see, e. g., \cite{a16,a17}). In the following I will analyze in brief the approach of Omn\`es which, at least in my opinion, represents the first fundamental and broadly elaborated recovering strategy \cite{a18}.

Omn\`es states that an object is "presumably associated with a family of states and a privileged category of observables." It is not clear which concept of state is used in this presumption, and the author gives neither a precise definition of "observable" nor of "privileged category", but now, based on this quite wobbly ground, the author considers macroscopic objects and assumes "that a supposedly well-defined object {\bf O} is associated" with a Hilbert space $\mathcal{H}_{\bf O}$. Then {\bf O} shall be characterized by a set of self-adjoint operators on $\mathcal{H}_{\bf O}$ which represent the so-called collective observables as, e. g., the momentum of the object's center of mass.

The crucial point is the connection between a quantal and the corresponding classical property. A quantal property is obtained by forming the trace over the product of a projector and the statistical operator representing the ensemble in question. A classical property, however, is established directly if the $n$-tuple ($q,p$) of spatial and momentum coordinates is element of a certain phase space cell. So the essential task is to construct a connection between projection operators and phase space cells.

Omn\`es starts from the one-dimensional and time-independent Schr\"odinger equation and applies the approximation of Wentzel, Kramers, and Brillouin, i. e., he makes the ansatz
\begin{equation}
\Psi = \mathrm{e}^{\frac{\mathrm{i}}{\hbar} \, \sigma(x)}
\end{equation}
and tries to solve the emerging differential equation by expanding the unknown function $\sigma(x)$ in powers of $\hbar/\mathrm{i}$. If this power series is stopped after the first-order term, one obtains the following three equations by sorting according to powers of $\hbar$:
\begin{eqnarray}
- \left( \frac{d \sigma_0}{dx} \right)^2 + p^2 &=& 0 \\ \hbar \, \left( \mathrm{i} \, \frac{d^2 \sigma_0}{dx^2} - \frac{2}{\mathrm{i}} \, \frac{d \sigma_0}{dx} \, \frac{d \sigma_1}{dx}\right) &=& 0 \\ \hbar^2 \, \left( \frac{d^2 \sigma_1}{dx^2} + \left( \frac{d \sigma_1}{dx} \right)^2 \right) &=& 0
\end{eqnarray}
(50) represents a correction of the order of $\hbar^2$ and is omitted. By use of the two surviving equations an approximate $\Psi$ can be calculated. Confinement of the motion of the object to an interval $[a,b]$ leads to boundary conditions which give rise to an equation similar to the old Bohr-Sommerfeld quantization condition:
\begin{equation}
\int_a^b p(x) \, dx = \left(N - \frac{1}{2}\right) \frac{h}{2}
\end{equation}
The integral is equal to $p_m (b - a)$ where $p_m \in \{p(x)|a \le x \le b\}$. Therefore it can be identified with the area $\mu(\mathrm{C})$ of a rectangular cell C of the phase space, and (51) may be written as $\mu(\mathrm{C}) = h \, (N - 1/2)/2 := h \, N(\mathrm{C})$. If Planck's constant is interpreted as a unit area of this phase space, and if we take into account that due to Heisenberg's uncertainty relation every quantum state requires a certain minimum area, then $N(\mathrm{C})$ is a measure of the number of states in C. In the case of $n$ degrees of freedom we have $\mu(\mathrm{C}) = h^n N(\mathrm{C})$.

In general, however, C will not be of rectangular shape which causes the main problem of finding a projector associated with a cell. It can be solved only if C is covered completely by a set of nonintersecting boxes, each with the same volume. This approximation allows to define a projector $\hat{P}(\mathrm{C})$, but different coverings will give different projectors. So it seems reasonable not to look for a unique projector but rather for a whole family of more or less equivalent projectors. Two projectors $\hat{P}_1(\mathrm{C})$ and $\hat{P}_2(\mathrm{C})$ are considered equivalent if $\mathrm{Tr} \, |\hat{P}_1(\mathrm{C}) - \hat{P}_2(\mathrm{C})|$, the so-called trace norm of their difference, is equal to zero. If, on the other hand, $\mathrm{Tr} \, |\hat{P}_1(\mathrm{C}) - \hat{P}_2(\mathrm{C})| \le \eta$, where $\eta$ is a yet undefined parameter, then the two projectors are called similar. By a kind ov averaging over said similar projectors a so-called quasi-projector $\hat{F}(\mathrm{C})$ can be generated which satisfies the following relations:
\begin{eqnarray}
\mathrm{Tr}(\hat{F}) &\approx& N(\mathrm{C}) \\ Tr \, |\hat{F} - \hat{F}^2| &\approx&  N(\mathrm{C}) f(\eta)
\end{eqnarray}
In this way quasi-projectors can be associated with cells, but it should have become clear from the derivation of (52) and (53) that this recovering strategy rests on a lot of significant approximations, and it seems very improbable that one can get rid of them. Moreover, the character of the whole process is quite arbitrarily, more an attempt to \emph{simulate} ClM instead of actually recovering it, and in consequence it is hard to believe that there exists a version of quantum mechanics which is really isomorphic to ClM in the strict sense of the word.

There are of course numerous other attempts to bridge the gap between QM and ClM. Ghirardi and coworkers have developed a "unified dynamics for microscopic and macroscopic systems" \cite{a19,a20} which rests on the idea to add correction terms to the Schrödinger equation "which are totally negligible at the level of one, two, or even 100 particles but play a major role when the number of particles involved becomes macroscopic (say of the order of $10^{23}$)." The activity of said correction terms is governed by two parameters, $\alpha$ and $\lambda$, which, in the end, decide about the existence of superpositions in dependence of the object size. In this way a \emph{transition} from QM to ClM and vice versa becomes possible, but one problem of this concept of a unified dynamics is the total arbitrariness in fixing exact values for $\alpha$ and $\lambda$: The requirement that a hydrogen atom shall be described quantum-mechanically but a grain of sand classically gives a very rough domain only. Note, moreover, that Ghirardi's approach is rather a combination of two independent theories on the same level than a recovering strategy!

It should be mentioned that, in contrast to Ghirardi, Yoneda et al. \cite{a21} tried to explain the appearance of classical features \emph{within} QM by introducing a coarse-grained position operator which, however, deserves an extra dimension. But again two parameters, $\Gamma$ and $\gamma$, play a crucial role, and their values are system-dependent as well.
\section{The approach of Nelson}
Some decades ago Nelson succeeded to derive Schr\"odinger's equation from Newtonian mechanics by use of the \emph{hypothesis} that every particle of mass $m$ is subject to a Brownian motion with diffusion coefficient $\hbar/(2m)$ and no friction \cite{a22}. This derivation, however, suffers from two critical items:
\begin{itemize}
\item The introduction of Planck's quantum of action as a part of the diffusion coefficient is completely arbitrarily. Why does it govern the Brownian motion mentioned above?
\item It implies the existence of zero-point vacuum fluctuations which exert stochastic forces that lead to the Brownian motion of the particles. Why should the \emph{classical} vacuum play this active role?
\end{itemize}
At least the assumption that there \emph{are} vacuum fluctuations is corroborated by the Casimir effect. In contrast, however, to Boyer's opinion \cite{a23} this effect does not result from fluctuations of a classical vacuum but, according to quantum electrodynamics, from fluctuations of the virtual particles forming the vacuum field. Therefore the existence of the Casimir effect does not rule out a \emph{totally} empty \emph{classical} vacuum. The work of Puthoff \cite{a24} could be used as another putative argument in favor of classical vacuum fluctuations. Closer inspection, however, reveals that Puthoff investigates the origin of zero-point fluctuations on the basis of charged point particles interacting with an electromagnetic zero-point radiation with a spectral-energy density corresponding to an average energy of $\hbar \omega /2$ per normal mode. In this way QM enters the discussion of the fluctuations' origin already at the very first stage! So, strictly speaking, also Puthoff deals with a quantum vacuum and not with the classical one which is the basis for Nelson's derivation.

But let us for the moment assume that also the classical vacuum is not empty but contains a field with zero-point fluctuations. In this way the energy $E$ of a particle would no longer be conserved along its trajectory in a conservative force field. Fritsche and Haugk have correlated the energy deviation $\Delta E$ with the energy-time uncertainty relation $\Delta E \, \Delta t \ge \hbar/2$ \cite{a25}. The use of this relation is essential for their derivation of the Schr\"odinger equation from Newtonian mechanics. But by use of $\Delta E \, \Delta t \ge \hbar/2$, which is of solely quantum mechanical origin, they have already introduced a crucial non-classical feature into ClM. So it is no surprise that Fritsche and Haugk achieve the desired result, but one should ask: What is the nutritive value of the whole attempt?

To sum up: Does it seem possible to derive Schr\"odinger's equation from ClM \emph{without} adding anything at all to ClM? By sure not, because then QM would be a mere subset of ClM and in consequence the predictions of QM and ClM might not differ from one another. The opposite, however, is the case.
\section{Conclusions}
In the preceding sections it has been pointed out that
\begin{itemize}
\item the transition $\hbar \rightarrow 0$, which is the essential step in the process to gain the Hamilton-Jacobi equation from Schr\"odinger's equation (or its analogue with statistical operators), not in any case removes the additional quantum potential created in the process of this derivation \cite{a9},
\item Ehrenfest's theorem is neither a necessary nor a sufficient condition to characterize the quantal$\rightarrow$classical transition \cite{a11},
\item the superposition principle has no counterpart in ClM, and
\item the application of QM to the solar system leads to absurd results.
\end{itemize}
Furthermore, typical constituents of the classical macroscopic world do not show up the typical quantum behavior which manifests
\begin{itemize}
\item in the wave-particle duality observed by means of the famous double-slit experiment and
\item in the EPR correlations.
\end{itemize}
Last but not least it is quite doubtful whether present attempts to recover ClM from QM will succeed. The inverse trials to obtain QM from ClM always need to add \emph{non}-classical features to ClM in advance, thereby plugging the desired result in its own derivation.

Summing up all these aspects we are compelled to agree with G\"unther Ludwig \cite{a26}:
\begin{itemize}
\item The assumption that "the Newtonian mechanics of our planetary system is only an approximation of a 'quantum theory' of the planetary system" is wrong. "Also the description of tennis balls via a classical mechanics is not an approximation of a 'quantum mechanics' for tennis balls $\ldots$ since it is physically impossible to make experiments with tennis balls similar to the interference experiments with electrons."
\item "$\ldots$ quantum mechanics is not more comprehensive than classical theories."
\end{itemize}
R\'{e}sum\'{e}: QM and ClM occupy different realms of nature. Therefore a wavefuntion of the universe does not exist.
\begin{appendix}
\newcounter{saveeqn}
\newcommand{\alpheqn}{\setcounter{saveeqn}{1}}
\setcounter{equation}{0}
\renewcommand{\theequation}{\mbox{\Alph{saveeqn}-\arabic{equation}}}
\alpheqn
\section{Stability of the solar system under a time-independent perturbation}
Coupling to a gravitational field causes the decay of the initial wavefunction according to
\begin{displaymath}
\Psi_n \rightarrow \sum_{l=1}^{\infty} c_l(t) \Psi_l.
\end{displaymath}
Let the coupling start at a time $t_0 = 0$. In first order perturbation theory the coefficient of the ground state $\Psi_1$ is then given by
\begin{equation}
c_1(t) = - \frac{i}{\hbar} \, \int_0^t \langle \psi_1 | \hat{H}^{\prime}(R,t^{\prime}) | \psi_n \rangle \times \mathrm{e}^{i \omega_{1n} t^{\prime}} \, d t^{\prime}
\end{equation}
with $\omega_{1n} = (E_1 - E_n)/\hbar$. The external field shall be generated by the planet jupiter which, for the sake of simplicity, is kept at a fix position $r_{SJ} = 7.88 \times 10^{11}$ m apart from the sun, i. e., we assume that the perturbation is time constant. In a two-dimensional polar coordinate system with the sun in its origin and jupiter
on the $x$-axis the perturbation operator is given by
\begin{equation}
\hat{H}^{\prime}(R) = - \frac{G M_1 M_3}{R}
\end{equation}
with
\begin{equation}
R = \sqrt{r_{SJ}^2 + r^2 - 2 r_{SJ} r \cos\phi}
\end{equation}
where $r$ and $\phi$ are the coordinates of the earth and $M_3$ is jupiter's mass. By use of the power series expansion of the square root the perturbation operator can be simplified to
\begin{equation}
\hat{H}^{\prime}(r,\phi) \approx - \underbrace{\frac{G M_1 M_3}{r_{SJ}}}_{:= \beta} \left( 1 - \frac{r^2}{2 r_{SJ}^2} + \frac{r}{r_{SJ}} \, \cos\phi \right).
\end{equation}
For the matrix element of the perturbation operator we then obtain
\begin{eqnarray}
\langle \psi_1 | \hat{H}^{\prime}(r,\phi) | \psi_n \rangle_{r,\phi} &=& - \, \beta \underbrace{\langle \psi_1 | \psi_n \rangle}_{= 0} \nonumber \\ & & + \, \frac{\beta}{2 r_{SJ}^2} \, \langle \psi_1 | r^2 | \psi_n \rangle_{r,\phi} \nonumber \\ & & - \, \frac{\beta}{r_{SJ}} \, \langle \psi_1 | r \, cos\phi | \psi_n \rangle_{r,\phi}.
\end{eqnarray}
Now we have at least two cases to consider. $\psi_1$, the wavefunction of the ground state, is obviously of $s$-type, i. e., $l = 0$. But which $l$-dependence shall we attribute to our $\psi_n$ with $n = 10^{74}$?

{\bf Case 1:} Let us assume first that also $\psi_n$ is an $s$-type wavefunction. This assumption corresponds to the selection rule $\Delta l = 0$ for the $\psi_n \rightarrow \psi_1$ transition. In this case the last integral in (A-5) reduces according to
\begin{equation}
 \langle \psi_1 | r \, cos\phi | \psi_n \rangle_{r,\phi} = \underbrace{\int_0^{2 \pi} \cos\phi \, d\phi}_{= 0} \, \times \, \langle \psi_1 | r | \psi_n \rangle_r
\end{equation}
so that
\begin{equation}
\langle \psi_1 | \hat{H}^{\prime}(r,\phi) | \psi_n \rangle_{r,\phi} = \frac{\beta}{2 r_{SJ}^2} \, \langle \psi_1 | r^2 | \psi_n \rangle_r.
\end{equation}
\begin{equation}
\Rightarrow c_1(t) = \underbrace{- \, \frac{\beta}{2 (E_1 - E_n) r_{SJ}^2}}_{:= f} \: \langle \psi_1 | r^2 | \psi_n \rangle_r \, \left( \mathrm{e}^{i \omega_{1n} t} - 1 \right)
\end{equation}
The value of $f$ amounts to $4.57 \times 10^{-177}$ m$^{-2}$.

We now look for the time $t_d$ necessary for the decay of our excited state to the ground state. It is given by the condition
\begin{equation}
|c_1(t_d)|^2 = f^2 \, \langle \ldots \rangle^2 \: (2 - 2 \cos(\omega_{1n} t_d)) = 1.
\end{equation}
\begin{equation}
\Rightarrow t_d = \frac{1}{\omega_{1n}} \arccos \left(1 - \frac{1}{2 f^2 \langle \ldots \rangle^2} \right)
\end{equation}
In order to be well defined the argument of the arccos function must belong to the interval $[-1,+1]$. This leads to the condition $f^2 \langle \ldots \rangle^2 \ge 1$. Unfortunately, due to the fact that the wavefunction $\Psi_n$ contains a Laguerre polynomial of the order $10^{74}$, it is impossible to evaluate the integral $\langle \psi_1 | r^2 | \psi_n \rangle_r$. So we can only say that, \emph{if} the arccos argument meets the above requirement, \emph{then} the \emph{maximum} decay time of the solar system will be equal to $\pi / \omega_{1n} = 5.09 \times 10^{-148}$ s in veraciously striking contrast to our daily experience.

{\bf Case 2:} Let us now assume that the relaxation of the solar system is governed by another selection rule, say, $|\Delta l| = 1$. In this case $\psi_n$ must be a $p$-type wavefunction given by
\begin{equation}
\psi_n = R_n(r) \times \underbrace{P_1(\cos\phi)}_{= \cos\phi}
\end{equation}
where the normalization constant is included in the radial part. Instead of (A-5) we now obtain
\begin{eqnarray}
\langle \psi_1 | \hat{H}^{\prime}(r,\phi) | \psi_n \rangle_{r,\phi} &=& \frac{\beta}{2 r_{SJ}^2} \, \underbrace{\int_0^{2 \pi} \cos\phi \, d \phi}_{= 0} \, \times \, \langle \psi_1 | r^2 | R_n \rangle_r \nonumber \\ & & - \, \frac{\beta}{r_{SJ}} \, \underbrace{\int_0^{2 \pi} \cos^2 \phi \, d \phi}_{= \pi} \, \times \, \langle \psi_1 | r | R_n \rangle_r.
\end{eqnarray}
\begin{eqnarray}
\Rightarrow c_1(t) &=& \underbrace{\frac{\pi \beta}{(E_1 - E_n) r_{SJ}}}_{:= g} \: \langle \psi_1 | r | \psi_n \rangle_r \, \left( \mathrm{e}^{i \omega_{1n} t} - 1 \right) \\ \Rightarrow t_d &=& \frac{1}{\omega_{1n}} \arccos \left(1 - \frac{1}{2 g^2 \langle \ldots \rangle^2} \right)
\end{eqnarray}
$g$ attains a value of $- 2.25 \times 10^{-164}$ m$^{-1}$, but $g$ appears in the arccos argument only so that this case does lead to the \emph{same} result as case 1. Note that this is valid for $|\Delta l| = 2$ as well.
\end{appendix}

\end{document}